\documentclass[traditabstract]{aa} 
\usepackage[english,greek]{babel}
\usepackage{graphicx}
\usepackage{txfonts}
\usepackage{wasysym}
\newcommand{\gtap}{\mathrel{\hbox{\rlap{\lower.55ex \hbox {$\sim$}}
                   \kern-.3em \raise.4ex \hbox{$>$}}}}
\newcommand{\ltap}{\mathrel{\hbox{\rlap{\lower.55ex \hbox {$\sim$}}
                   \kern-.3em \raise.4ex \hbox{$<$}}}}

\begin{document}
\selectlanguage{english}
  \title{On the disruption of pulsar and X-ray binaries in globular clusters}

  \author{Frank Verbunt\inst{1,2} \and Paulo C.C.\ Freire\inst{3}}

  \institute{Department of Astrophysics/IMAPP, Radboud University Nijmegen, PO Box 9010,
    6500 GL Nijmegen, The Netherlands; \email{F.Verbunt@astro.ru.nl}
  \and SRON Netherlands Institute for Space Research, Utrecht
  \and Max-Planck-Institut f\"ur  Radioastronomie, auf dem H\"ugel 69,
    Bonn, Germany; \email{pfreire@mpifr-bonn.mpg.de}}

  \date{Accepted \today}

  \abstract{The stellar encounter rate $\Gamma$ has been shown to
  be strongly correlated with the number of X-ray binaries in
  clusters and also to the number of radio pulsars. However, the pulsar
  populations in different clusters show remarkably different
  characteristics: in some GCs the population is dominated by binary
  systems, in others by single pulsars and exotic systems that result
  from exchange encounters.
  In this paper, we describe a second dynamical parameter
  for globular clusters, the encounter rate for a single binary, $\gamma$.
  We find that this parameter provides a good characterization of
  the differences
  between the pulsar populations of different globular clusters.
  The higher $\gamma$ is for any particular globular cluster
  the more isolated pulsars and products of exchange interactions
  are observed. Furthermore, we also find that slow and ``young'' pulsars
  are found almost exclusively in clusters with a high $\gamma$;
  this suggests that these kinds of objects are formed by the
  disruption of X-ray binaries, thus halting the recycling of a
  previously dead neutron star.
  We discuss the implications of this for the nature of young pulsars
  and for the formation of neutron stars in globular clusters.}


  \maketitle

\section{Introduction}
\label{s:intro}

  Millisecond pulsars (MSPs), also known as recycled pulsars, are
  the progeny of X-ray binaries in which a neutron star (NS) accretes
  matter from a companion (as reviewed by e.g.\ Phinney \&\ Kulkarni
  1994, Stairs 2004). Since X-ray binaries are overabundant in
  globular clusters with respect to the galactic disk (as reviewed by
  e.g.\ Heinke 2011), it was predicted that  millisecond pulsars
  would be present in globular clusters  (Fabian et al.\ 1983).  This
  prediction has been abundantly verified by the observations.

Close encounters between stars and/or binaries are much more frequent
in globular clusters than in the galactic disk (Hills 1975), and this
is thought to be the cause of the overabundance of X-ray binaries in
globular clusters (Clark 1975). \nocite{hil75}\nocite{cla75}
In early publications, three types of
encounter were suggested to bring a neutron star into a binary: direct
collision with a giant, tidal capture, and exchange encounters
(Sutantyo 1975, Fabian et al.\ 1975, Hills 1976). We will refer to
these as primary encounters. The numbers of neutron stars
were estimated from an assumed initial mass function, e.g. Salpeter's,
the parameters of which were determined from the number counts of the
main-sequence stars, adapting the method devised (for white dwarfs) by
Tinsley (1968); \nocite{sut75}\nocite{tin68}
or by proxy from the central surface brightness with
an assumed mass-to-light ratio.  The rate of each type of encounter
roughly scales with the encounter number $\Gamma$ (Verbunt \&\ Hut
1987):
\begin{equation}
\Gamma \propto  {{\rho_c}^2{r_c}^3\over v} =  K  {\rho_c}^{1.5} {r_c}^2 
\label{e:gamma}\end{equation}
where $\rho_c$ and $r_c$ are the density and radius of the cluster
core, $v$ is the velocity dispersion, and $K$ a constant. 
From the virial theorem $v\propto \sqrt{\rho_c}\,r_c$.
The encounter number $\Gamma$ has been fairly successful in describing
the numbers
of X-ray binaries, both bright and quiescent (Pooley et al.\ 2006,
Heinke et al.\ 2006), and the numbers of MSPs (Johnston et al.\ 1992).
When the number of radio pulsars is estimated from a luminosity
function, however,  results for a correlation with the encounter
number are mixed (Hui et al.\ 2010, cf.\ Bagchi et al. 2011).
\nocite{hct10}\nocite{blc11}\nocite{bhsg13}
Bahramian et al.\ (2013) make an effort to compute more
accurate values for $\Gamma$ from photometric cluster profiles,
which enables them to treat core-collapsed clusters in the same
manner as other clusters,
and conclude that the correlations between numbers of X-ray
and radio sources with $\Gamma$ are, in fact, strong.
Interestingly, they find that core-collapsed clusters may contain
{\em fewer} X-ray sources that predicted by their $\gamma$ values.

The successful use of $\Gamma$ should not obscure the fact that there
are many aspects of the formation of X-ray binaries and their possible
evolution into MSPs which $\Gamma$ does not describe. Examples are the
differences between clusters in the slope of the initial mass
function, in the properties of the binary population, in the fraction
of stars lost from the cluster, and in the retention fraction of
neutron stars.  In addition, the importance of encounters {\em
  subsequent} to the primary encounters is not taken into account by
$\Gamma$. We will refer to such subsequent encounters as secondary
encounters.  To investigate the importance of these aspects, detailed
numerical calculations have been made by e.g., Sigurdsson \&\ Phinney
(1995) and Ivanova et al.\ (2008).

Sigurdssson \&\ Phinney (1995) assumed that 10\%\ of the neutron stars
are born with velocity low enough to remain in the cluster. They
consider stationary, albeit evolved, globular clusters. Their
calculations show that collisions with low-mass stars are important
only in low-density clusters; for clusters with increasing central
density the encounters are dominated more and more by the more massive
stars, including the neutron stars, that fully dominate the encounters
in a model with a central density similar to that of M\,15. Secondary
encounters in high-density clusters ionize low-mass binary pulsars to
produce a population of single pulsars. Other single neutron stars and
neutron stars with a companion of very low mass are produced by direct
collisions in high-density clusters between main-sequence stars and
neutron stars (as originally suggested by Krolik et al.\ 1984). Such
direct collisions occur also in three-body encounters between a
neutron star and a binary and in four-body encounters -- i.e.\ the
encounter between two binaries (e.g.\ Bacon et al.\ 1996, Fregeau et
al.\ 2004.) In the model for the high-density cluster discussed 
by Sigurdsson \&\ Phinney (1995), for example, a neutron star involved
in a three-body encounter merges with both other stars involved.
If this neutron star accretes enough material from the remnants
of the merger stars, it may become a single millisecond pulsar.

Ivanova et al.\ (2008) allow the clusters to evolve. Importantly,
Ivanova et al.\ assume that neutron stars due to core collapse are
born with high velocities (based on Hobbs et al.\ 2005) , so that only
a tiny fraction of them is retained. The neutron star population in
globular clusters in their calculations are dominated by those formed
in electron-capture supernovae, which are assumed to be born with
velocities ten times lower (see Podsiadlowski et al.\
2004).\nocite{plp+04}  Electron
capture may occur in single stars, but may also be induced by
accretion in a merger of two white dwarfs, or in a binary. In the
latter case the low velocity enhances the probability that the newly
formed neutron star remain in the binary.  A result of these new
channels for the formation of a neutron star is an increased
importance and bewildering variety of secondary encounters, as
graphically illustrated in Fig.\,1 of Ivanova et al.\ (2008).  Ivanova
et al.\ (2008) find that most low-mass X-ray binaries are formed by
such accretion-induced collapse in low-density clusters, whereas in
high-density clusters binary exchange, collisions with giants and
tidal capture together also contribute significantly (some 50\%) to
the formation of low-mass X-ray binaries.  Neutron stars formed by
accretion-induced collapse probably are not spun-up to MSPs in the
same binary, but by companions obtained after secondary encounters. In
high-density clusters single MSPs arise from neutron stars that merge
with a main-sequence donor, or by ionization in an exchange encounter.
The fraction of single MSPs increases with the density of the cluster,
reaching 70\%\ in the densest models considered.

The description in the previous two paragraphs cannot do full justice
to the complexity of the numerical calculations.  Even so, major
uncertainties remain, as may be illustrated by two examples. The life
time of a low-mass X-ray binaries is determined among others by the
loss of mass and angular momentum from the binary, both of which are
essentialy unknown. The susceptibility of a binary to secondary
encounters scales directly with its life time. An uncertainty by a
factor ten in the life time of the binary translates into a major
uncertainty in the importance for its evolution of secondary
encounters. The velocity distribution of newly born neutron stars must
have a low-velocity component, but the size and form of this component
is not much constrained by the observations due to significant
measurement uncertainties (Hartman 1997, Brisken et al.\ 2002).
To what extent this low-velocity component is dominated by neutron
stars formed via electron-capture is equally unknown.
The importance of this question is illustrated if we juxtapose
the conclusion from Sigurdsson \&\ Phinney ``{\em We do not find
it necessary to invoke AIC to account for any globular cluster MSP
formation}'' to the one by Ivanova et al.\ (2008) ``{\em To reproduce
the empirically derived formation rate of LMXBs we must assume that NSs
can be formed via electron-capture supernovae}''.

In this our paper we take a closer look at the observed properties
of MSPs in globular clusters, to see whether they provide clues to
the importance and nature of secondary encounters. We do so on the
basis of the single-binary encounter rate $\gamma$.
In Section\,\ref{s:disruption} we
explain this number and show that it correlates well
with the ratio of the number of single pulsars to the
number of pulsars in a binary (Section\,\ref{s:isolated})
and that binary systems that resulted from
exchange encounters occur mostly in GCs with high $\gamma$
(Section\,\ref{s:exchange}).
In Section\,\ref{s:periods} we analyse the spin period
distribution for several GCs, and discuss the occurrence
of slow (Section\,\ref{s:slow}) and young, high-B field
(Section\,\ref{s:young}) pulsars in different clusters.
Unlike the previous section, which just discusses
binary disruption in general, this section discusses
objects that are more likely to be produced by
the disruption of X-ray binaries and the resulting
truncation of the spin-up and field ``burial'' processes.
A discussion of our results is given in
Section\,\ref{s:discussion}, in particular regarding
the implications for the origin of NSs in GCs, their formation
rates and their relation to the young pulsars
observed in GCs.

\section{The single-binary encounter rate}
\label{s:disruption}

Those pulsars in GCs that are recycled by accretion from a
binary companion, are
in binaries with the donor remnants, in particular low-mass
WDs. As explained above, their formation rate scales roughly
with the encounter rate $\Gamma$ of the cluster.
However, once a binary exists, it may undergo subsequent encounters.
Analogously to Eq.\,\ref{e:gamma} one may estimate the encounter
rate for one object, e.g.\ a binary, as (Verbunt 2003):\nocite{ver03}
\begin{equation}
\gamma   \propto { \rho_c\over v} = C {\sqrt{\rho_c}\over r_c}
\label{e:sgamma}\end{equation}
If one considers the cross section $A$ for a specific
type of encounter, the expected life time of the
binary until the encounter may be written
\begin{equation}
\tau = {1\over\gamma}
\label{e:tau}\end{equation}

In clusters with a high value of $\gamma$,
binaries are prone to encounters
that increase their average orbital eccentricities and
increase their chances of being disrupted, leading either
to the formation of single pulsars or the replacement of
the companion of a recycled neutron star with another star
(an ``exchange encounter'').
We investigate the results of such encounters in the following
subsections.

\begin{table}
\caption{\label{t:clusters}
Structural parameters and numbers of single and binary pulsars
 of globular clusters with three or more pulsars \label{t:numbers}}
\begin{tabular}{clcccrrr}
 & cluster & log $r_c$ & log $\rho_c$  & $\gamma$ & $N_s$
& $N_b$ \\
 &  & (pc) & ($L_\odot$\,pc$^{-3}$) & ($\gamma_\mathrm{M\,4}$)\\
\hline
a & NGC\,6522    &    $-$0.95c &     5.48 &    55.1  &   3 &   0 \\
b & NGC\,6624    &    $-$0.86c &     5.30 &    36.4  &   5 &   1 \\
c & NGC\,6517    &    $-$0.73c &     5.29 &    26.8  &   3 &   1 \\
d &  NGC\,6752    &    $-$0.70c &     5.04 &    18.8  &   4 &   1 \\
e &  NGC\,6440    &    $-$0.46 &     5.24 &    13.5  &   3 &   3 \\
f &  Terzan 5      &    $-$0.49 &     5.14 &    13.0  &  16 &  18 \\
g & NGC\,6441    &    $-$0.36 &     5.26 &    10.9  &   2 &   2 \\
h & NGC\,6266/M\,62    &    $-$0.36 &     5.16 &     9.8  &   0 &   6 \\
i & NGC\,7078/M\,15    &    $-$0.37c &     5.05 &     8.9  &   7 &   1 \\
j & NGC\,6626/M\,28    &    $-$0.42 &     4.86 &     7.9  &   4 &   8 \\
k & NGC\,104/47\,Tuc  &    $-$0.33 &     4.88 &     6.6  &   8 &  15 \\
l & NGC\,5904/M\,5    &    $-$0.02 &     3.88 &     1.0  &   1 &   4 \\
m & NGC\,5272/M\,3    &     0.04 &     3.57 &     0.6  &   0 &   4 \\
n & NGC\,6205/M\,13    &     0.11 &     3.55 &     0.5  &   2 &   3 \\

\end{tabular}
\tablefoot{Core radius $r_c$ and central density $\rho_c$ from Harris (1996,
update of 2010), with ``c''s indicating a core-collapsed cluster.
The disruption rate $\gamma$ is scaled on the value for M\,4, from
Eq.\,\ref{e:sgamma}. Numbers $N_s$ of single pulsars and $N_b$ of pulsars in
binaries from {\tt www.naic.edu/}$\sim${\tt pfreire/GCpsr.html}
(retrieved 7 Sep 2012)}
\end{table}

\subsection{Fractions of Isolated pulsars in Globular Clusters}
\label{s:isolated}

\begin{figure}
\centerline{\includegraphics[angle=0,width=\columnwidth]{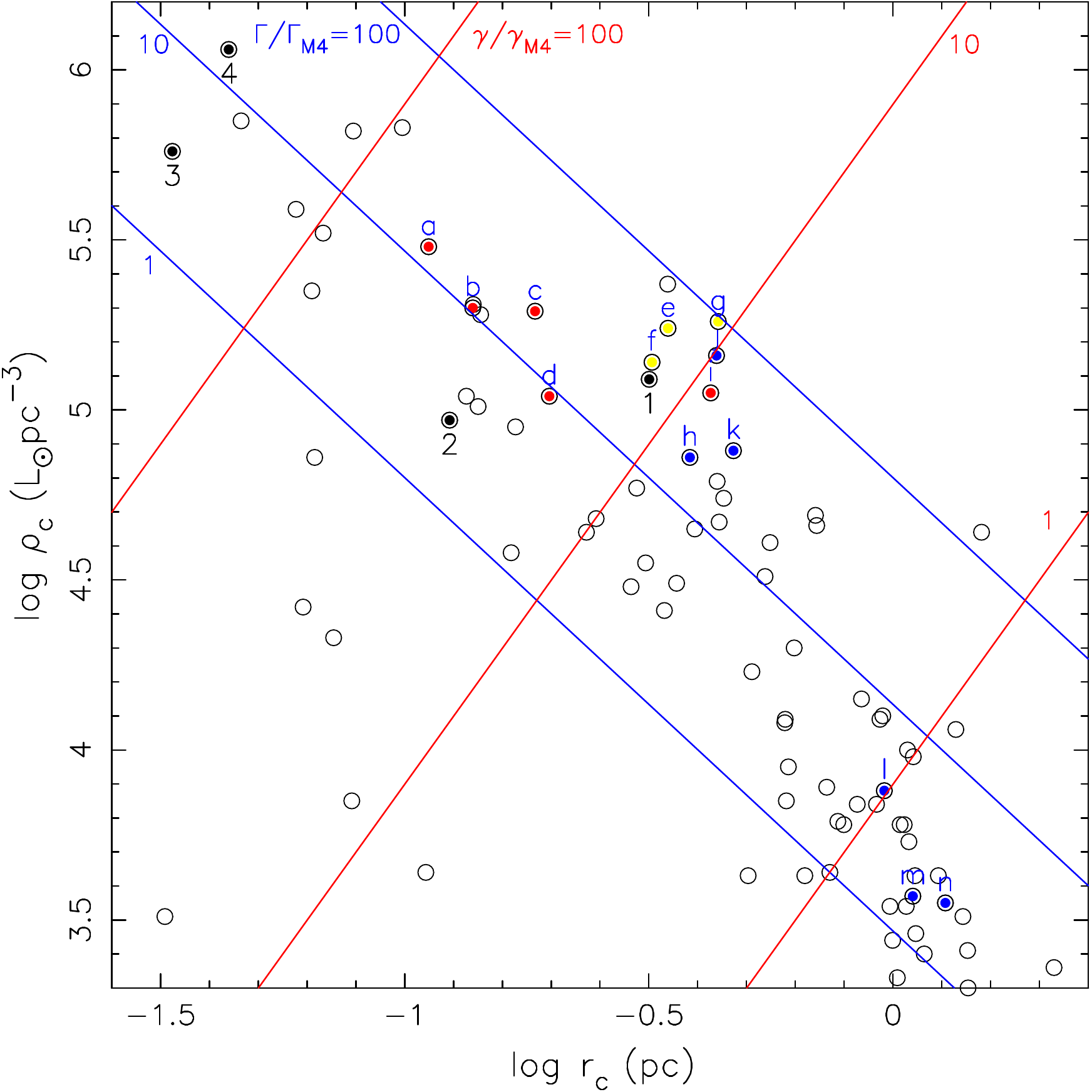}}

\caption{Central density $\rho_c$ as a function of core radius $r_c$
for clusters with $\rho_c>2000\,L_\odot/\mathrm{pc}^3$
(Harris (1996, update of 2010), with lines of constant $\Gamma$
(blue) and constant $\gamma$ (red). Letter labels refer to clusters in
Table\,\ref{t:numbers}, coloured red if $N_s>N_b$, blue if $N_b>N_s$,
and yellow if $N_b\simeq N_s$. Clusters with products of secondary interactions
include i) and g), and -- indicated in black, with numerical labels --
1: NGC~1851, 2; NGC~6342, 3: NGC~6397 and 4: NGC~6544.
   \label{f:rhorc}}
 \end{figure}

In Figure\,\ref{f:rhorc} we show the central density as a function of
core radius for clusters with $\rho_c>2000\, L_\odot/\mathrm{pc}^3$.
In the figure we indicate lines of constant collision number $\Gamma$
and lines of constant single-binary disruption rate $\gamma$,
normalizing both on the values for the GC M\,4 to eliminate the
constants $K$ and $C$ from Eqs.\,\ref{e:gamma} and \ref{e:sgamma}.
The central density is expressed in solar luminosity per cubic parsec,
and in plotting $\Gamma$ and $\gamma$ we implicity assume that the
mass density scales linearly with the luminosity density.  In clusters
with high $\gamma$, we expect a relatively short life time of a binary
before it is disrupted or undergoing an exchange (depending also on
its size, i.e.\ semi-major axis), and thus expect a relatively high
fraction of the pulsars to be released from the binary in which it was
spun up, and thus a high fraction of single pulsars (Sigurdsson \&\
Phinney 1995). Single pulsars are also formed by merging with a
main-sequence donor, according to Ivanova et al.\ (2008), who also
predict a larger fraction of single pulsars in high-density clusters.

To see whether these predictions are supported by observations, we
list in Table\,\ref{t:numbers} some parameters for all globular
clusters in which three or more pulsars have been detected.  In
addition to the central density, core radius and (normalized)
single-binary disruption rate we list the numbers of single pulsars
$N_s$, and of pulsars in a binary $N_b$.  We have
ordered the clusters on $\gamma$, and it is seen that the clusters in
which $N_s>N_b$ are at the top of the Table, and those in which
$N_b>N_s$ at the bottom.

For most individual clusters the numbers are too small for significant
conclusions. To illustrate this we test the null hypothesis that
the probability for a pulsar to be single is the same for all clusters
and equal to the observed fraction of single pulsars for all clusters
listed in Table\,\ref{t:numbers} combined:  $p=$58/(58+67)=0.464.
For each cluster we compute the probability $Q_s$ that the observed
number or more of single pulsars are found; and the probability $Q_b$
that the observed number or more of pulsars in a binary are found,
given the above value of $p$ and the total number of pulsars observed
in that cluster, with use of binomial statistics. 
Only two significant ($>2\sigma$) deviations from the null
hypothesis are found: $Q_s$ = 0.022 for NGC\,7078,  and $Q_b$ = 0.024 for NGC\,6266.

Collectively, however, the difference between the
clusters with highest and lowest values of $\gamma$ are very
significant. For the four clusters with the highest $\gamma$ together,
$Q_s$ is 0.0015; for the four (three) clusters with the lowest $\gamma$ together
$Q_b$ is 0.0029 (0.015).
This is further illustrated in Fig.\,\ref{f:rhorc} where the clusters
with $N_s\,>\,N_b$ are indicated in red, those with $N_b\,>\,N_s$ in blue,
and those with $N_s\,\simeq \, N_b$ in yellow.

Given that many of the caveats made in Section\,\ref{s:intro} 
for the use of $\Gamma$ apply equally to the use of $\gamma$,
the overall correlation of the fraction of single pulsars with
$\gamma$ is remarkable. NGC~7078 is an exception. 
We wish to make two remarks concerning this. First, NGC\,7078
is the only cluster, to our knowledge, for which a pulsar search was
made by performing a very long FFT after stacking data from several
days. This method is only sensitive for single pulsars, and found
several very faint pulsars (Anderson 1993). Thus the number of single
pulsars is biased in NGC\,7078. (In contrast, a search on
stacked data sets of up to 6 hrs of 47\,Tuc did not discover new
pulsars; Knight 2007.)  \nocite{kni07}
The second remark is that NGC\,7078 illustrates another
complication in the use of $\Gamma$ and $\gamma$: the uncertain
values of the cluster parameters. Many values for the core
radius have been published. Those based on HST imaging
yield small estimates of $r_c$:  2\farcs2 (Lauer et al.\
1991)\nocite{lhf+91}, $<$1\arcsec\ (Guhathakurta et al.\ 1996)\nocite{gypb96},
1\farcs78 (Dull et al.\ 1997)\nocite{dcl+97}, $<1$\farcs3
(Sosin \& King 1997)\nocite{sk97}, 4\farcs07
(Noyola \& Gebhardt 2006)\nocite{ng06}; such values would yield
a much higher values $\gamma$ and smaller value of $\Gamma$
for this GC. The compilation we use,
Harris (1996, updated 2010)\nocite{har96}, gives 0\farcm14, the
the average of the values in Trager et al. (1995)\nocite{tkd95}
and in Lehmann \& Scholz (1997)\nocite{ls97}, both of which are
based on ground-based imaging. 

Clusters that show no clear flattening in the central light
distribution, as measured with ground-based telescopes, have
traditionally been classified as `core-collapsed'. Since
clusters just before core collapse may observationally be
very similar to clusters just after core collapse (e.g.\ Trenti 
et al.\ 2010)\nocite{tvp10},  the classification `core-collapsed'
is perhaps best interpreted as reflecting a very small core,
and thus probably a high $\gamma$. In this view, it is remarkable
that {\it all} clusters in Table\,\ref{t:numbers} that have more single
than binary pulsars are classified as core-collapsed.

Figure\,\ref{f:gamp} shows the pulse period as a
function of $\gamma$ for all GCs, including those
in which only one or two pulsars are found. This plot shows
60 single pulsars and 84 binary pulsars.
In clusters with low $\gamma$
($\gamma<3\gamma_\mathrm{M4}$)
we find 5 single pulsars and 20 pulsars in a binary; 
taking the probability for a pulsar to be single as 60/(60+84), 
we find $Q_b=0.02$. In clusters with high $\gamma$ 
($\gamma>16\gamma_\mathrm{M4}$) we find 15 single pulsars
and 9 pulsars in a binary; for these clusters $Q_s=0.03$.
Our conclusion that high-$\gamma$ clusters are more likely to
hold single pulsars thus holds when we consider all known
pulsars in clusters.

\begin{figure*}
\centerline{\includegraphics[angle=0,width=15cm]{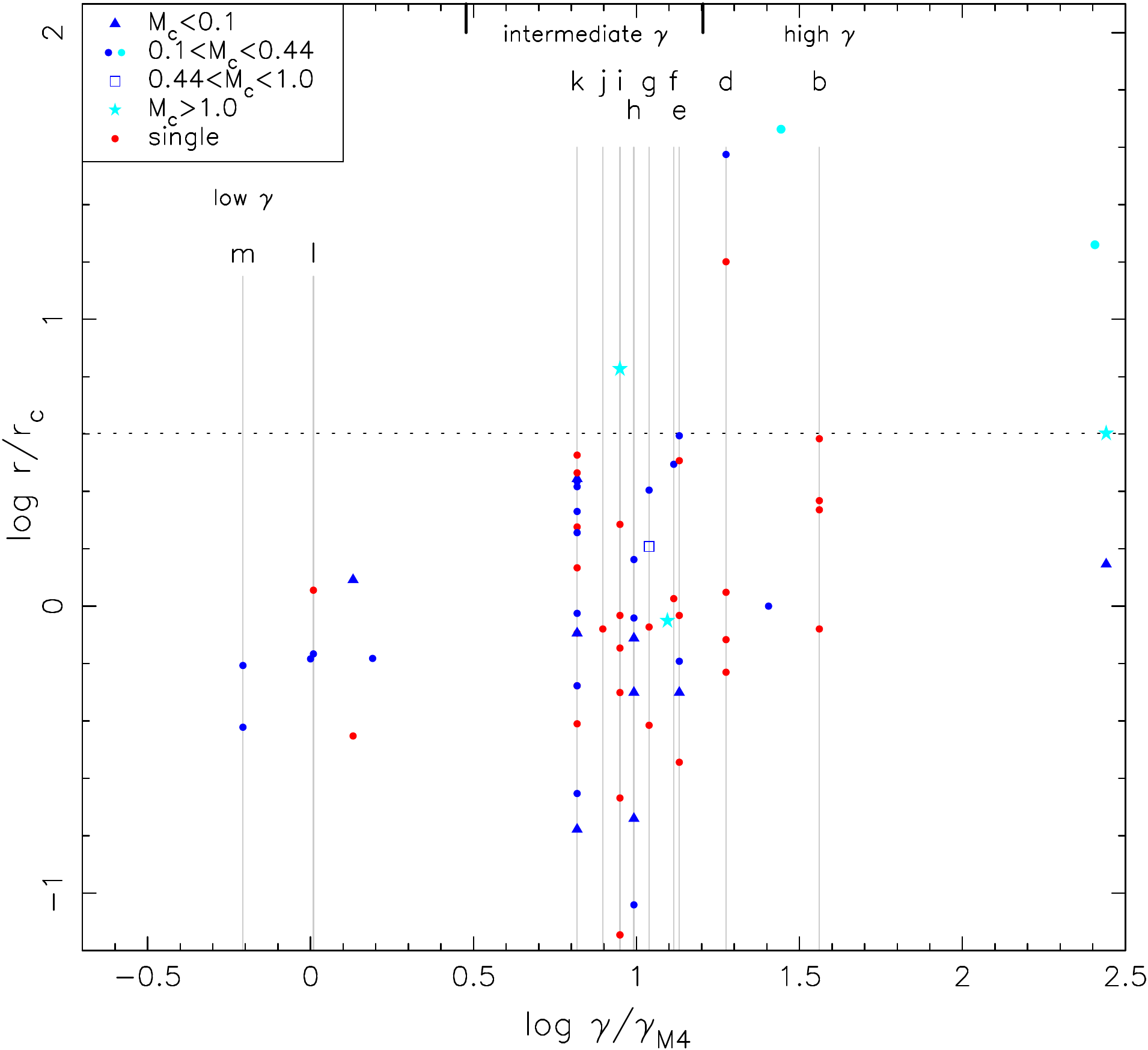}}
\caption{Angular offsets of the pulsars with published timing solutions
from the centres of their clusters, in core radii. The
larger the encounter rate per binary $\gamma$ the larger are the odds
of finding pulsars far from the centres of their clusters.
Some of the clusters are labelled as in Table\,\ref{t:numbers}, colours
are used to discriminate isolated and binary pulsars, and symbols to
indicate the companion mass. Binaries created by exchange products are
shown light blue. The dashed line indicates a distance of 4 core radii.\label{f:rgamma}}
 \end{figure*}

\subsection{Dynamical state of the pulsar populations in globular clusters}
\label{s:dynamics}

Whenever a pulsar is involved in an exchange encounter, it
suffers a kick, no matter whether it finds itself in a binary or in
isolation. This can unbind the pulsar (or binary pulsar) from the
cluster, or give it a highly elliptical orbit about its centre
(Phinney 1992 and references therein). If it still remains
bound by the cluster, then with time it will interact with other
cluster stars (generally lighter) and gradually lose kinetic energy to
them (a phenomenon known as ``dynamical friction''), therefore
returning to the core of the cluster, where most pulsars are found in
a state often described as of relaxed equilibrium --- which means that
the average kinetic energy of all stellar species there are very
similar, and therefore their radial distributions have no tendency to
change.

Indeed, the latter state is what is found for all globular
clusters with a low $\gamma$, where all pulsars are
within 4 core radii of the center (see Fig.~\ref{f:rgamma}).
The pulsar distribution in 47\,Tuc, which is thought to be relaxed,
is also limited to within this radius.
As $\gamma$ increases, we start
seeing pulsars at much larger distances from the centres
of their clusters. These have been involved in strong dynamical
interactions in the relatively recent past and had yet no time
to sink back to the core of the cluster through dynamical
friction. The exclusive presence of distant pulsars in high-$\gamma$
clusters is consistent with the idea that these large positional
offsets really are the result of binary disruption and
exchange interactions.

\subsection{Eccentricities}

Initially circular binaries may be made eccentric by a mild encounter
with a third star, and the eccentricity of non-circular binaries 
generally increases when mild encounters occur (e.g.\ Rasio \&\
Heggie 1995)\nocite{hr95}. Indeed, the dissolution of a binary
discussed in Sect.\,\ref{s:isolated} may be considered as the inducing of
an eccentricity larger than unity.  The probability of an encounter
scales linearly with the semi-major axis of the binary, or in terms of
Eq.\,\ref{e:sgamma}: $A\propto a$.  The induced eccentricity has a
stronger dependence on the semi-major axis, or more precisely on the
ratio of the semi-major axis and the closest approach $r_p$ of the
third body; in particular for $r_p\gg a$ it scales with
$(a/r_p)^{5/2}$ for a circular orbit and with $(a/r_p)^{3/2}$ for an
already eccentric orbit (Heggie \&\ Rasio 1996).

Collisions between a neutron star and a giant may directly produce
an eccentric binary of a neutron star and white dwarf, with orbital
periods less than a day (Ivanova et al.\ 2005).

\begin{figure}
\centerline{\includegraphics[angle=0,width=\columnwidth]{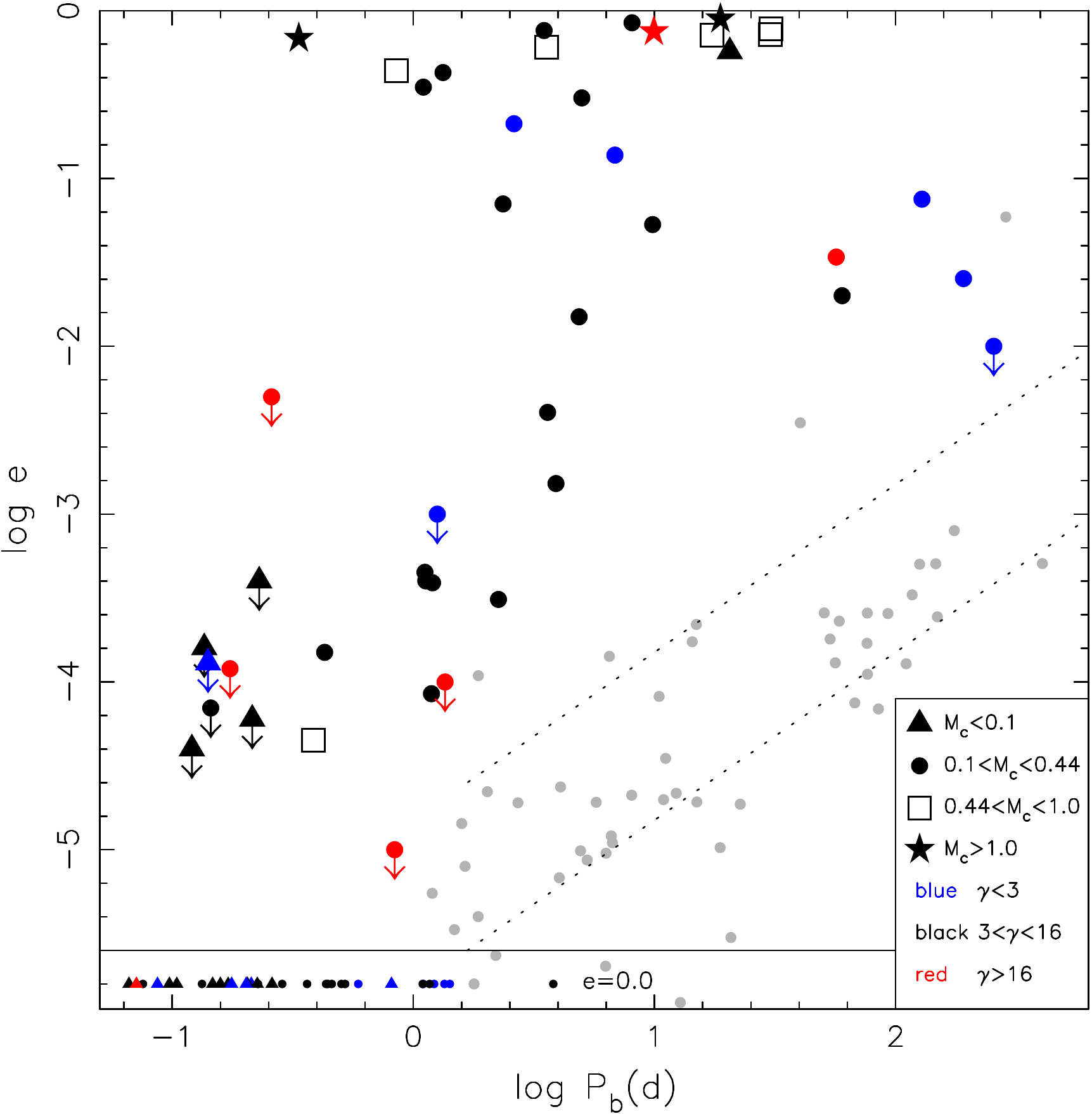}}
\caption{Eccentricity as a function of orbital period for binary radio
pulsars in globular clusters. Colours are used to indicate $\gamma$
of the cluster in which the pulsar resides, symbols to indicate
the companion mass. For the outcome of recycling by a (sub)giant donor
one expects $P_b\gtap1$\,d and $0.1<M_c/M_\odot<0.44$; such systems 
in the Galactic Disk are shown in gray, and the expectation values for
the eccentricity according to Eq.\,\ref{e:ecc} and ten times this
value, are shown as dotted lines.
Eccentricies 0.0, for which no upper limits have
been specified, are shown at the bottom (with smaller symbols
to avoid overlap). Data from
{\tt http://www.naic.edu/}$\sim${\tt pfreire/GCpsr.html} 
\label{f:pbe}}
 \end{figure}

In Fig.\,\ref{f:pbe} we plot the orbital eccentricities of cluster
pulsars as a function of their binary periods.
Pulsars in binaries with orbital periods less than a day may have
evolved from binaries with main-sequence companions, or from binaries
with white dwarf donors that were acquired by the neutron star as it
collided with a giant.  For most of these only upper limits to the
eccentricity are known. This agrees with the fact that short-period
binaries have small probabilities of engaging in a secondary encounter.

The orbital periods ($\gtap1$\,d) and companions masses (0.1 - 0.44
$M_\odot$) of many of the systems shown in Fig.\,\ref{f:pbe} indicate
an origin from a binary with a (sub)giant donor. From the
fluctuation-dissipation theorem such binaries -- when unperturbed --
have expectation values for the eccentricity given by (Phinney
1992):\nocite{phi92}
\begin{equation}
<e^2>^{1/2} \simeq 1.5\times10^{-4} {P_b\over 100\,\mathrm{d}}
\label{e:ecc}\end{equation}
For binary radio pulsars in the galactic disk, shown in
Fig.\,\ref{f:pbe} as gray dots, this relation holds quite well.  In
globular clusters, the eccentricities of {\it all} binary radio
pulsars with companion masses 0.1 - 0.44 $M_\odot$ are larger, which
indicates that all of them have had eccentricity-enhancing secondary
encounters with third stars.

For binary pulsars in GCs with intermediate and high
$\gamma$ (shown black and red in the Figure) the orbits with periods
longer than about a day are concentrated in Fig.\,\ref{f:pbe} towards
higher eccentricities. These binaries may have
experienced several eccentricity-enhancing secondary encounters,
which indicates that encounters leading to eccentricities larger than 1,
i.e.\ unbinding the pulsar, are not uncommon.

Several of the pulsars in orbits with high eccentricities have
companions with masses $0.45\le M_c/M_\odot\ltap1.0$.  These
companions are possibly carbon-oxygen white dwarfs, evolved from
donors on the asymptotic giant branch. This evolution leaves a wide
binary, initially with very small eccentricity, but highly susceptible
to eccentricity-enhancing secondary encounters.

\subsection{Exchange encounters}
\label{s:exchange}

Carbon-oxygen white dwarfs or neutron stars can also become companions
to pulsars via an alternative route: having formed from a single star,
they subsequently replaced a lower-mass white-dwarf companion to the
pulsar in an exchange encounter.  Companions in GCs with masses $M_c\gtap
M_\odot$ certainly entered the binary with the neutron star in an
exchange encounter, irrespective of whether they are high-mass white
dwarfs or neutron stars.  They are indicated in Fig.\,\ref{f:pbe} with
a star. One of the striking features of this Figure is that all such massive
companions have very eccentric orbits, as one would expect from
exchange encounters.

We now discuss these products of exchange encounters.
PSR~J1807$-$2500B in NGC\,6544 (\cite{lfrj12}) has a massive ($M_c =
1.2064(20)\rm \, M_{\odot}$) companion, which is either a massive WD
or a light NS, in a highly eccentric ($e = 0.75$) orbit with a period
close to ten days. Whatever the companion is, its progenitor was too
massive and too short lived to recycle the pulsar to its present spin
period of 4.15 ms. 
This implies that the pulsar was recycled in a binary formed
in a primary encounter, and that its current companion exchanged
into the binary in a secondary encounter, in which the first
companion of the neutron star was expelled. 
As discussed above, the probability of a secondary encounter
does not depend on $\Gamma$, only on $\gamma$.
NGC\,6544 in fact has the highest single-binary encounter
rate, $\gamma=276\gamma_\mathrm{M4}$, of all clusters
containing a pulsar.

Another example of a secondary exchange interaction is PSR~B2127+11C,
in NGC\,7078 = M15 (\cite{agk+90}), a 30.5-ms pulsar in an a tight (8.0 hour),
eccentric ($e = 0.68$) orbit with a NS. Timing this pulsar, Prince et
al. (1991) derived a characteristic age of about 0.1 Gyr (this is little
affected by the acceleration of the binary in the potential of M15).
Although the
system superficially resembles the first binary pulsar, B1913+16 (see
Weisberg, Nice \& Taylor 2010 and references therein\nocite{wnt10}),
the progenitor of its massive companion could not have recycled the pulsar
0.1 Gyr ago --- at that time there were no such massive stars in the cluster.
The system must have therefore formed through one or more
exchange interactions (\cite{pakw91}). The pulsar is found quite far
from the cluster center, again consistent with an exchange
interaction and associated recoil (Section~\ref{s:dynamics}).
As discussed above, M15 has an intermediate value of $\gamma$
(see Table\,\ref{t:numbers}).

The third clear example of an exchange encounter is PSR~J0514$-$4002A,
in NGC~1851 (\cite{fgri04}); this is the most eccentric binary pulsar
in all GCs and it also has a massive companion (\cite{frg07}), which
again might be a massive WD or a very light NS (or even the core of a
giant star, which might explain the anomalous presence of diffuse gas
in the system). NGC\,1851 has an intermediate single-binary encounter
rate, $\gamma=12.4\gamma_\mathrm{M4}$.

An exchange encounter may also result in a pulsar acquiring a normal
main-sequence star companion. An example of this is another remarkable
binary, PSR~B1718$-$19 in NGC~6342 (\cite{lbhb93}).  The pulsar has a
relatively young characteristic age, with a spin period of 1 second,
and is orbited every 6.2 hours by a non-degenerate star
(\cite{kkk+00}). This secondary is likely to have replaced the
pulsar's previous companion in a secondary exchange encounter;
the system's position, far from the cluster center,
offers further support of this hypothesis (Section~\ref{s:dynamics}).
The short binary period
has enabled the non-degenerate star to reduce the eccentricity, which
is observed to be small: $e<0.005$\footnote{ 
An alternative explanation for PSR~B1718$-$19 is that the binary
is the result of a primary encounter, in which the current companion
lost some of its mass, of which only a small fraction was accreted
by the neutron star. In either case, it might be about to become
a new X-ray binary.}. NGC\,6342 has a single-binary
encounter rate $\gamma=28\gamma_\mathrm{M4}=28$.

A final example of this is PSR~J1740$-$5340 
(\cite{dlm+01,dpm+01}) in NGC~6397, the cluster
with the second highest $\gamma$ among all in the Galaxy. The pulsar has a
spin period of 3.65 ms; its companion is a bloated,
optically bright star (\cite{fpds01}). The system is
at least 20 core radii from the cluster centre, which again suggests
a large recoil produced by an exchange encounter.

The resulting pattern is very clear: only clusters with intermediate
or high encounter rates per binary ($\gamma$) host pulsar binaries
formed in secondary exchange encounters.

\section{Pulse period distributions and incidence of
young pulsars with strong magnetic fields}
\label{s:periods}

\begin{figure*}
\centerline{\includegraphics[angle=0,width=15cm]{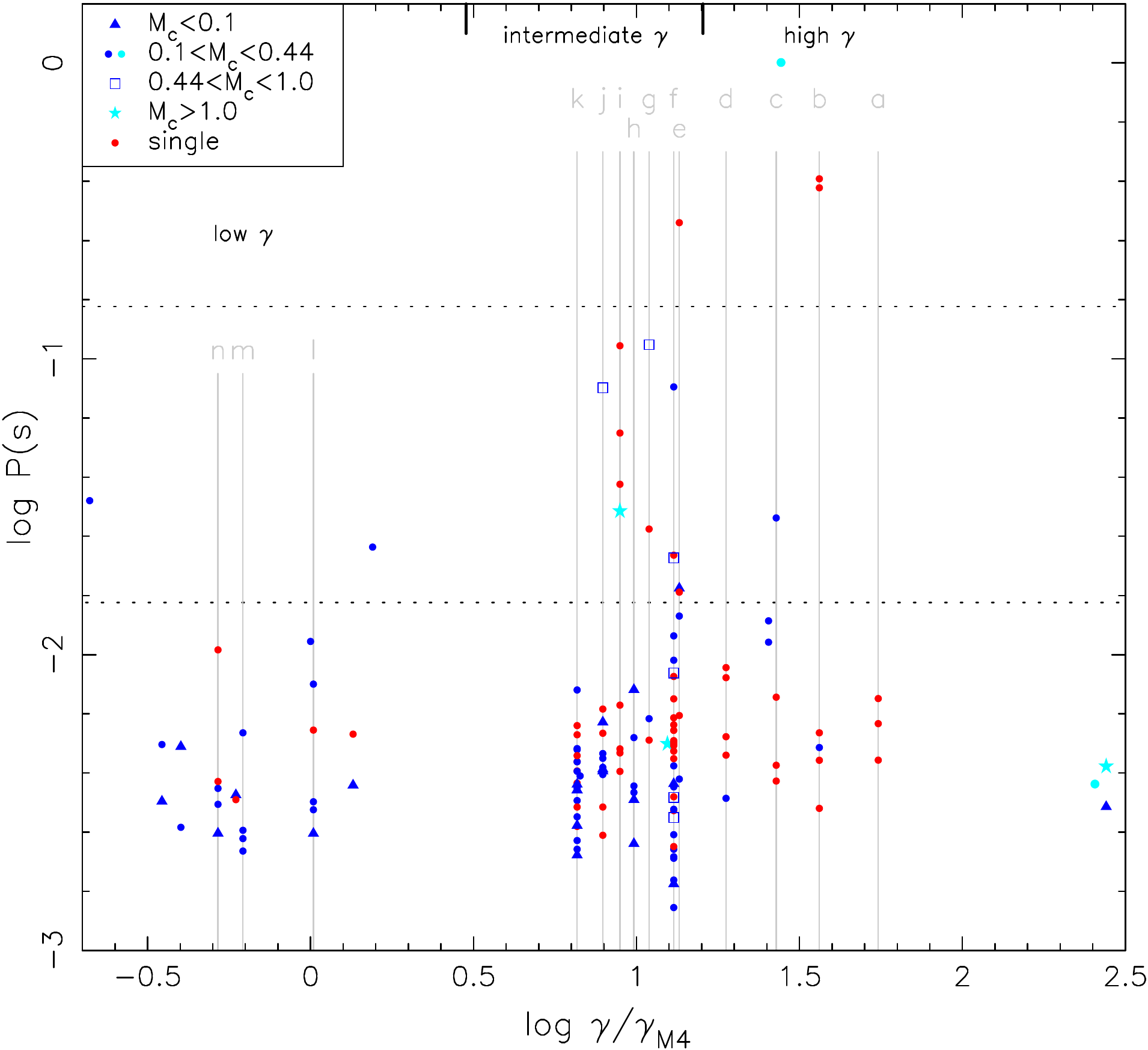}}
\caption{Pulse periods as function of single-binary disruption rate $\gamma$.
Some of the clusters are labelled as in Table\,\ref{t:numbers}, colours
are used to discriminate isolated and binary pulsars, and symbols to
indicate the companion mass. Binaries created by exchange products are
shown light blue. The horizontal dotted lines indicate pulse periods of
0.15 and 0.015\,s.
\label{f:gamp}}
\end{figure*}

A striking result from the previous Section is that {\em all} pulsar
binaries with $P_b\gtap 1$\,d have had an encounter that significantly
enhanced their eccentricity.
This encounter can have happened during the phase of mass transfer,
as the neutron star was being spun up, or after cessation of the mass
transfer. Three time scales are involved: the duration of the mass
transfer phase, the time passed since the cessation of
the mass transfer during which the neutron star
is a recycled radio pulsar, and the waiting time for an encounter
given by Eq.\,\ref{e:tau}.
In clusters with a very low $\gamma$, a binary is expected to
complete its mass-transfer phase and remain unperturbed; from Fig.\,\ref{f:pbe}
we see that no pulsars have been found in wide, unperturbed binaries, presumably
because clusters with very low central density have very low $\Gamma$
as well,  and thus are unlikely to harbour {\em any} recycled pulsar.
Clusters with a somewhat higher $\gamma$ will allow completion
of the mass-transfer phase for most systems, and induce the
eccentricity only afterwards.
In clusters with an even higher $\gamma$ the mass transfer is liable to
disruption, either permanently or temporarily, by a close encounter,
in particular in systems with wide orbits.
 
Can we discriminate between systems that had their encounter during or after the
mass-transfer phases? During the mass-transfer phase the neutron star
is being spun up and its magnetic field is being reduced.  Systems in
which mass transfer has been disrupted by an encounter may therefore
be expected to have, on average, longer pulse periods and
stronger magnetic fields. The characteristic age of the pulsars in such
systems reflects the time passed since the interaction, which 
may be much smaller than the age of the cluster.

\subsection{Pulse periods}
\label{s:slow}

In Fig.\,\ref{f:gamp} we display the spin periods
of the pulsars in GCs as a function of the single-binary
disruption rate $\gamma$.
It is apparent that the pulsar populations in
different GCs have different spin period distributions.
This is not caused by observational
biases. First, there is generally no bias against the detection of
slow pulsars, which could be detected for all GCs but are only
detected in some.
There are biases against the detection of fast pulsars,
but they have been largely overcome in recent years.
As an example, until recently the pulsars in Terzan 5 had the highest
DMs among all known MSPs ($\sim 240 \, \rm cm^{-3}\,pc$),
which precluded the
detection of fast MSPs. However, improvements in instrumentation
now allow the detection of pulsars with very fast spin periods
(\cite{rhs+05}), including the fastest known, PSR~J1748$-$2446ad
(\cite{hrs+06}), even for such high DMs.

This means that we can now have a very good idea of the
spin period distribution in GCs, particularly when several
GCs are surveyed with consistent observational setups.
Hessels et al. (2007)\nocite{hrs+07} observed many of the
GCs within the Arecibo declination range with a very high
time resolution.
Although they discovered fast-spinning pulsars in
M3, M5, M13, M71 and NGC~6749 they did not find any new
fast-spinning pulsars in M15, which has significantly smaller
dispersive smearing than M71 and NGC~6749. Such bright,
fast-spinning pulsars would have eluded previous 430~MHz
surveys surveys but should be easily detectable in this survey.
This means that the relatively small fraction of fast-spinning
pulsars in M15 is a real effect, not a selection effect.

In Fig.\,\ref{f:gamp} we see that short pulse periods are
observed at all values of $\gamma$. However, the pulsars with pulse
periods in the intermediate range of 0.015\,s -- 0.15\,s are
observed mostly at intermediate values of $\gamma$, and those with
pulse periods longer than 0.15\,s at intermediate to high values
of $\gamma$. Of the 19 pulsars with pulse periods above 0.015\,s, 9
are single.
Two pulsars in this period range are in clusters with low $\gamma$.
The distribution of intermediate pulse periods forms a bridge between
the shortest pulse periods and the longest pulse periods, which
suggests that the latter are the tail of a continuous distribution
rather than a separate group. 

Figure\,\ref{f:gamp} suggests the following scheme for the
spin-up of pulsars in GCs. In clusters with low $\gamma$ X-ray
binaries of all periods mostly live long enough for their NSs
to be spun up to MSPs.
In clusters with intermediate $\gamma$ the binaries with short periods
live long enough for full spin-up to occur, but some binaries with long
orbital periods can be disrupted or changed in such a way that mass
transfer stops, leaving pulsars with intermediate pulse periods,
either isolated (in the first case) or in a binary (in the second case). 
Finally, in clusters with high
$\gamma$ only short-period binaries live long
enough for spin-up to short periods to be completed, and long-period
binaries tend to be disrupted before much spin-up has occurred, hence
the presence of pulsars with large spin periods there.
The prevalence of single
pulsars in high-$\gamma$ clusters indicates that even binaries with
extended periods of mass transfer are eventually disrupted, either
during the X-ray binary phase or later as binary MSPs, as discussed
in Section\,\ref{s:disruption}.

This scenario explains very well the result of Hessels et al. (2007)
that there are many fast pulsars in M3, M5 and M13, and slower pulsars
in M15: the former clusters have low $\gamma$. However, $\gamma$
cannot be the whole story.  The GC M\,53 has
$\gamma=0.21\gamma_\mathrm{M4}$, but the pulsar in it has a
pulse period of 0.033\,s.  The orbital period of this pulsar is the
longest one observed in a GC, 255.8\,d, with a companion
mass of 0.35\,$M_\odot$.  When mass transfer started in the progenitor
binary, the giant was already close to the tip of the giant branch,
and rapidly expanding.  Its envelope may have been transferred quickly
and at super-Eddington rate, leading to inefficient spin-up. Similar
cases in the galactic disk are J0407+1607 and J2016+1948, which
have orbital periods of a few hundred days and spin periods of
tens of ms (Lorimer et al.\ 2005, Navarro et al.\ 2003).\nocite{lxf+05}\nocite{naf03}

Also, when Freire et al. (2008) observed NGC\,6440/6441 with exactly
the same system used to find the fast-spinning pulsars in Terzan~5
they found distinctively slower populations in NGC\,6440/6441, even
though these clusters have a slightly lower dispersive smearing and no
detectable scattering, i.e., pulsars as fast and bright as
PSR~J1748$-$2446ad should be clearly detectable.  In a
Kolmogorov-Smirnoff test, they found $<$ 2\% probability for the
pulsars populations of NGC~6440 and 6441 being drawn from the same
sample as the pulsars in Terzan~5. The $\gamma$'s for these three
clusters are very similar, which implies that the assumptions made in
simplifying the estimate of $\gamma$ (similar initial mass function,
similar escape velocity, etc., see Sect.\,\ref{s:intro}) may be violated.

\begin{figure}
\centerline{\includegraphics[angle=0,width=\columnwidth]{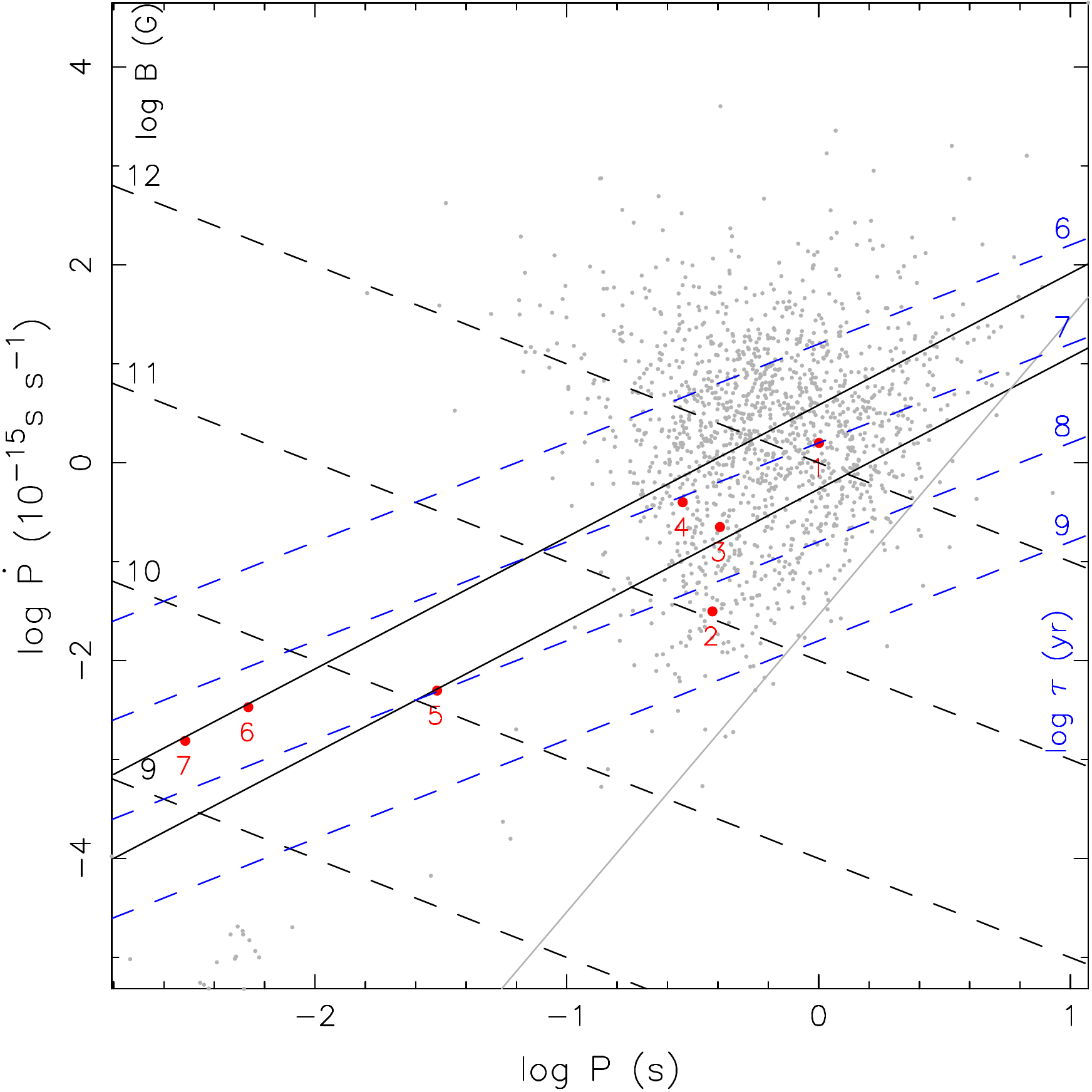}}
\caption{Pulse period derivatives as function of pulse period for
GC pulsars with a strong magnetic field and/or short characteristic
age. Numbers according to Table\,\ref{t:special}.
Grey points indicate single pulsars in the Galactic Disc, the grey
solid line the death-line,
the black solid lines are the spin-up line according to
Eq.\,\ref{e:spinup} and a line shifted by a factor 7 in $\dot P$.
Dashed lines of constant magnetic field (black) and constant age 
(blue) are marked with their respective values.
\label{f:ppd}}
\end{figure}

\subsection{Magnetic fields and characteristic ages}
\label{s:young}

\begin{table}
\caption{\label{t:special} Pulsars with small characteristic ages in
  globular clusters}
\begin{tabular}{cllcccc}
& NGC: PSR & $P$ & $\dot P$ & $B$  &
$\tau_c$  \\
 & & (s) & (10$^{-15}$) & (10$^{12}$\,G) &(10$^{10}$\,yr)\\
\hline
1 & 6342: B1718$-$19 &   1.0040 &    1.59000 &    1.263 &    0.001 \\
2 & 6624: B1820$-$30B &   0.3786 &    0.03150 &    0.109 &    0.019 \\
3 & 6624: J1823$-$3021C &   0.4059 &    0.22400 &    0.302 &    0.003 \\
4 & 6440: B1745$-$20 &   0.2886 &    0.39933 &    0.339 &    0.001 \\
5 & 7078: B2127+11C &   0.0305 &    0.00499 &    0.012 &    0.010 \\
6 &6624: B1820$-$30A &   0.0054 &    0.00338 &    0.004 &    0.003 \\
7 &6626: B1821$-$24A &   0.0031 &    0.00155 &    0.002 &    0.003 \\
\end{tabular}
\tablefoot{Because the age of the globular clusters in our galaxy is
  close to 10$^{10}$\,yr, the last column may also be read as giving
  the characteristic age as fraction of the cluster age. References
  for these pulsars are given in Sect.\,\ref{s:young}.
}
\end{table}

The magnetic field strength and characteristic age of a pulsar are
estimated from the product and the ratio, respectively, of its spin
period $P$ and its derivative, $\dot{P}$.  The problem in GCs is that
pulsars accelerate in their gravitational fields. The line-of-sight
component of this acceleration, $a$, introduces a contribution to the
the observed $\dot{P}_{\rm obs}$ which is of the order of $a P/c$. For
most MSPs, this is of the same order of magnitude or larger than their
instrinsic spin-down $\dot{P}$, which implies that for most pulsars in
GCs we cannot estimate the magnetic fields. However, for
some pulsars in GCs $\dot{P}$ is so large (and positive) that it
cannot be attributed to the cluster acceleration $a$. We list those
that have strong magnetic fields and/or short characteristic ages in
Table\,\ref{t:special}, and show them in Fig.\,\ref{f:ppd}.

The four objects with the longest pulse periods in
Table\,\ref{t:special} are PSR~B1718$-$19 (\cite{lbhb93}),
PSR~B1820$-$30B (\cite{bbl+94}), J1823$-$3021C (\cite{lfrj12}) and
B1745$-$20 (\cite{lmd96}). In the $P$ vs. $\dot P$ diagram these
pulsars are located in the same area as the single pulsars in the
galactic disk, with strong magnetic fields and young characteristic
ages, compared to most pulsars in GCs. This has led to the suggestion
that these NSs were formed recently.  We note, however, that the
magnetic fields of three of these four pulsars are at the low end of the
distribution for young pulsars in the galactic disk, where the typical
magnetic field is of the order of $10^{12}$\,G, and that all four are
in GCs with intermediate or high values of $\gamma$. 
This suggests that they have undergone some
loss of field strength due to accretion in a short-lived X-ray binary, 
and that their relatively high remaining magnetic-field strengths and long
periods are due to disruption of the X-ray binary by a close
encounter and consequent interruption of the spin-up process.
Alternatively, these neutron stars were spun up by a limited amount 
of material that they collected in a direct collision with a
main-sequence star or giant.

In both scenarios, the characteristic age reflects the time passed since the
encounter -- which indeed may be much less than the cluster age.
We already noted in Sect.\,\ref{s:exchange} that PSR~B1718$-$19
may also be a product of an exchange encounter.
The strong magnetic fields and long periods suggest that
only a small amount of mass was accreted by the neutron star
before accretion stopped.
 
The three other pulsars in Table\,\ref{t:special}, B2127+11C,
B1820$-$30A (\cite{bbl+94,faa+11}) and B1821$-$24A (\cite{lbm+87,fbtg88,jgc+13}),
have lower magnetic fields but very small characteristic ages.  This
has led to the suggestion that these NSs were also formed recently.
Again, we suggest that these PSRs are not young neutron stars, but
recently recycled old neutron stars. As discussed in
Sect.\,\ref{s:exchange}, the B2127+11C system must be the result
of an exchange encounter, possibly the same that disrupted the
X-ray binary that spun up the pulsar.

The hypothesis that the young pulsars are recently recycled old
neutron stars implies that they must be below the spin-up line 
in the $P$ vs $\dot P$ diagram. This line indicates the shortest
period that can be reached by spin-up due to accretion 
at the Eddington limit onto a neutron star (see Pringle 
\&\ Rees 1972)\nocite{pr72}
\begin{equation}
P_{su}\mathrm{(s)}\simeq 1.3(B_{12})^{6/7} \Leftrightarrow
P_{su}\mathrm{(s)} \simeq 1.6 \left(10^{15}\dot P\right)^{3/4}
\label{e:spinup}\end{equation}
where we substitute $(B_{12})^2\simeq10^{15}P\dot P$. 
We plot this spin-up line in Fig.\,\ref{f:ppd}. The exact
values of the constants in Eq.\,\ref{e:spinup} depend on details
that are not well understood and are therefore uncertain. Indeed,
the constants may be different for different evolutionary histories
(Tauris et al.\ 2012).\nocite{tau12} Therefore,  we also
plot an alternative spin-up line for which $\dot P$ is a factor 7 higher at 
the same $P$.

If we opt for the line according to Eq.\,\ref{e:spinup}, we find that
we must accept not only some pulsars with $B>10^{11}$\,G but also
two -- B1820$-$30A and B1821$-$24A -- with rather lower
magnetic fields as young, rather than recycled neutron stars.
This means that we must not only accept that neutron stars
are still being formed in GCs, but also that they
are formed at least in part with rather weak magnetic fields and very
high spin period\footnote{These unusual objects would have
to be forming in rather large numbers as well. As discussed
in Freire et al. (2011),
NGC~6624A and M28A are about 100 times less numerous than the
normal MSPs in GCs, but they will be visible as MSPs
for a time that is  $\sim$ 100 times shorter as well. This implies
that  both groups must be forming at similar rates.}.

On the other hand, if we opt for the higher spin-up line, all
apparently young pulsars in GCs may be old neutron stars
recently spun up in now disrupted X-ray binaries. This is more
consistent with the fact that they only appear in GCs with 
intermediate or high $\gamma$.

\section{Discussion}
\label{s:discussion}

In the previous Sections we have assumed that the single-binary
disruption rate $\gamma$ is the factor which determines the fraction
of single pulsars, and -- together with the binary period -- the
degree in which the spin-up process can be completed. Sigurdsson
  \&\ Phinney (1995) and Ivanova et al.\ (2008) discuss the results of
  their calculations mainly in terms of the central density of the
  cluster.
From Table\,\ref{t:numbers} and Fig.\,\ref{f:rhorc} it is
evident that for the clusters under discussion there is an
almost perfect correlation between $\gamma$ and $\rho_c$, and an
equally perfect anti-correlation between $\gamma$ and $r_c$. Thus we
cannot, on the basis of the current evidence, decide from observations
which of the three is the determining factor. Our choice for $\gamma$
is on the basis of theory. It may be noted that $\rho_c$ varies over
two orders of magnitude for the clusters in Table\,\ref{t:numbers},
and $r_c$ over one order of magnitude in anticorrelation with
$\rho_c$. Since $\gamma\propto \sqrt{\rho_c}/r_c$, the variations in
central density and core radius contribute in approximately equal
measure to the variation in $\gamma$.  As may be seen in
Figure\,\ref{f:rhorc} we can exclude that the collision number
$\Gamma$ determines the fraction of single pulsars, which clusters
harbour the products of exchange encounters, or even which clusters
have slow pulsars and pulsars with high magnetic fields.

We have already noted that all clusters in Table\,\ref{t:clusters}
with more single than binary pulsars are core-collapsed. In this
context it is interesting to note a difference between $\Gamma$ and
$\gamma$ when these numbers are integrated over the cluster history.
Semi-analytic models for the cluster evolution (Lynden-Bell \&\
Eggleton 1980, Goodman 1984) show that the central density
scales with core radius as $\rho_c\propto {r_c}^{-\alpha}$, with
$\alpha\simeq 2.22$, and that the time that the core spends at radius
$r_c$ scales as $t_{rc}\propto v^3/\rho_c$. The number of
encounters at each radius thus scales as
\begin{equation}
\Gamma t_{rc} \propto {{\rho_c}^2{r_c}^3\over
v}{v^3\over\rho_c} \propto {r_c}^{5-2\alpha}
\label{e:gamint}\end{equation}
For $\alpha\simeq2.22$ the number of close encounters over the history of
a cluster is thus dominated by the times at large core radius.
We may write the number of encounters for a single object at given
radius as
\begin{equation}
\gamma t_{rc} \propto {\rho_c\over v}{v^3\over\rho_c} \propto {r_c}^{2-\alpha}
\label{e:lowgamint}\end{equation}
Thus the encounter rate for a single object peaks during core
collapse, {\em provided that the object participates in the collapse}.
We suggest that this may be an extra reason for the prevalence
of single pulsars in core-collapsed clusters.
It should be noted that the scaling relations we use here are for
drastically simplified cluster models -- in particular all stars
in these models have the same mass. Therefore it is a pity that
Ivanova et al.\ (2008) note that their computations cannot be applied to
core-collapsed clusters.

We note in passing that the enhanced number of secondary encounters,
as expressed in $\gamma$, in core-collapsed clusters may explain the
result by Bahramian et al.\ (2013), mentioned in our Introduction,
that core-collapsed clusters contain fewer X-ray binaries than
non-collapsed clusters with similar $\Gamma$.

The pulsars in Table\,\ref{t:special} have characteristic ages much
younger that the cluster age, and to explain this it has been
suggested that neutron stars are still being formed in globular
clusters.  To explain the observed magnetic fields of the young
pulsars in this way, one must accept that neutron stars formed via
electron capture (possibly induced by accretion) are born with fields
ranging from 10$^9$ to 10$^{12}$\,G, in marked contrast to the neutron
stars formed from core collapse, which lead to a magnetic field
distribution centered on 10$^{12}$\,G and ranging 10$^{11}$ to
10$^{13}$\,G.  The exclusive presence of young pulsars in GCs with
intermediate or high $\gamma$ indicates, as we highlighted above, that
the `young' pulsars are the result of mild accretion episodes, where
for instance the pulsar is a member of an X-ray binary for only a
short period until that binary gets disrupted, 
or alternatively accretion following a direct collision. 
This hypothesis
automatically explains why the young MSPs are single, or in the
case of PSR~B1718$-$19 and B2127+11C why they have currently
companions acquired through exchange encounters. It also explains the large range
in magnetic fields, which reflects the duration and rate of the mass
transfer before disruption occurred. In particular the contrast in
$\gamma$ between NGC\,6624 and 47\,Tuc may explain the remarkable 
fact that 3 of 6 pulsars in NGC\,6624 are apparently young, but {\em
  none} of the 23 in 47\,Tuc, 
 despite the fact that the latter cluster has a higher $\Gamma$.

Assuming that the pulsars in GCs are young, Boyles et
al. (2011)\nocite{blt+11} estimate that a birth rate of 1 every 10$^4$
yr is required to explain the observed number of young pulsars in the
GC system.  Since under our hypothesis a single neutron star can go through several
short-duration recycling and radio pulsar episodes, the required NS
formation and retention rate is potentially much smaller
and the characteristic age reflects the time since the most
recent encounter.

Note that secondary encounters are expected to occur in the
dense central core of the GC. This implies that the primary encounter
bringing the NS into the binary produce only a small systemic velocity
of the binary; or alternatively that enough time has passed for
dynamic friction to reduce higher systemic velocities which would put
the binary outside the central region. Fig.\,\ref{f:rgamma} shows that
such is indeed the case.

A final question is why we see so many isolated pulsars compared to
results of exchange encounters in high-$\gamma$ clusters. After all,
in an exchange encounter involving a NS, the most likely outcome is
the formation of a binary system containing the two most massive
components, i.e.\ most often the NS with a main-sequence companion.
The calculations of Sigurdsson \&\ Phinney (1995) point to two
  reasons in clusters with central densities as high as in M\,15:
  first, frequent encounters with neutron stars ionize low-mass pulsar
  binaries, and second, direct collisions, both in two- and three-body
  encounters, lead to single pulsars. The calculations by Ivanova et
  al. (2008) indicate two other channels as the most important ones:
  direct collisions of a neutron star with a main-sequence star leading
  to a merger, and exchange encounters releasing a pulsar from its
  binary. From the current observations we cannot discriminate between
  these possibilities.

\section{Conclusions and further work}

In this paper we use the encounter rate per binary in a globular cluster,
$\gamma$, to complement the total stellar encounter rate in
the cluster, $\Gamma$. The latter has been shown to be a reasonable indicator
of the number of X-ray binaries and pulsars in GCs; the former
provides a first indication of the characteristics of
the pulsar populations in GCs. The larger $\gamma$ is
the more disturbed are the MSP binaries: we see more secondary
exchange encounters, more isolated pulsars, and more
pulsars in the outer parts of the clusters, far from where one would expect
them to be given mass segregation. These large offsets result from the
recoils caused by binary interactions so recent that the pulsar has not yet
had time to sink back to the core by dynamical friction.
The correlation is not perfect: some clusters with similar $\Gamma$
and $\gamma$ have markedly different pulsar populations.
This reflects in part a lack of detailed and uniform characterization
of all the clusters, but might also be caused by different dynamical
histories, e.g.\ the occurrence of a core collapse.

We find that in general the placement of particular types of GC
objects in a $\log \rho_c - r_c$ (or $\Gamma-\gamma$) diagram of the
parent cluster can provide a useful diagnostic of the evolution of
that particular type of object.  An example of this is the young,
high-B field pulsars: these objects appear exclusively in GCs with
intermediate or high $\gamma$. This, together with the finding that
they are all near the same spin-up line suggests that they are
produced when X-ray binaries are disrupted or single-neutron stars
accrete a limited amount of debris after a direct collision with
a giant or main-sequence star.  Further evidence for this is the
large fraction of these systems that appear as members of exchange
products.  This greatly diminishes the large required rate of NS
formation in GCs calculated in Boyles et al. (2011) under the
assumption that these young pulsars are newly created NSs: in our
scenario, one single NS can go through several recycling and radio
pulsar episodes. Each recycling episode diminishes the magnetic field
of the NS, which for each pulsar incarnation appears successively
closer to the normal MSPs in the $P$-$\dot{P}$ diagram.

This interpretation can be put to the test in two ways. First,
any pulsar found in a GC well above the (upper) spin-up line in
Fig.\,\ref{f:ppd}, must be a genuinely young pulsars, rather
than an old pulsar recently involved in a secondary encounter.
Second, when more pulsars become known in low-$\gamma$
clusters, our interpretation predicts that
apparently young and long-period pulsars, as well
as highly eccentric pulsar binaries and systems clearly
the product of exchange encounters, should be very rare
among these.

Finally, this work suggests that exotic systems like
MSP-MSP or MSP-black hole binaries should appear
almost exclusively in medium to high-$\gamma$ clusters.
Therefore, future radio pulsar surveys for these objects
may be more successful in those globular clusters.

\begin{acknowledgements}
We thank the Lorentz Center in Leiden for organizing the
``Compact Binaries in Globular Clusters'' meeting, where the idea
for this paper started germinating. PF also thanks Ed van den
Heuvel for suggesting his attendance to said meeting.
We also thank Thomas Tauris for an interesting discussion regarding
the implication of our results for the origin of neutron stars in
globular clusters, and the referee Fred Rasio for helpful comments.
\end{acknowledgements}

\end{document}